# Imaging field-tuned quantum Hall broken-symmetry orders and quantum Hall conducting channel in charge-neutral graphene/WSe$_2$ heterostructure

Qi Zheng, Mo-Han Zhang, Ya-Ning Ren, Lin He*

Center for Advanced Quantum Studies, Department of Physics, Beijing Normal University, Beijing, 100875, People's Republic of China

*Correspondence and requests for materials should be addressed to Lin He (e-mail: helin@bnu.edu.cn).

**The zeroth Landau level (0LL) in graphene has emerged as a flat-band platform in which distinct many-body phases can be explored with unprecedented control by simply tuning the strength and/or direction of magnetic fields[1-22]. A rich set of quantum Hall ferromagnetic (QHFM) phases with different lattice-scale symmetry-breaking orders are predicted to be realized in high magnetic fields when the 0LL in graphene is half filled[1-8,13-16]. Here we report a field-tuned continuous quantum phase transition of different valley orderings in QHFM of charge-neutral graphene on insulating tungsten diselenide (WSe$_2$). The phase transition is clearly revealed by anomalous field-dependent energy gap in the half-filled 0LL. Via atomic resolution imaging of electronic wavefunctions during the phase transition, we observe microscopic signatures of field-tuned continuous-varied valley polarization and valley inversion, which are unexpected and beyond current theory predictions. Moreover, the topological quantum Hall conducting channel of the graphene is directly imaged when the substrate (WSe$_2$) introduces band bending of the 0LL.**

In flat band, the kinetic energy is quenched and electron-electron interactions can drive electrons to form exotic correlated phases to minimize the Coulomb interaction. For example, correlated insulator, superconductivity, orbital ferromagnetism, and quantum anomalous Hall effect are observed very recently in partially filled flat bands of magic-angle twisted bilayer graphene (MATBG)[23–26]. In graphene systems, introducing a perpendicular magnetic field is the simplest way to create the flat bands, *i.e.*, Landau levels (LLs), and realize exotic many-body phases (without involving the very accurate fine-tuning of the twist angle as in the MATBG)[1–22]. At half-filled zeroth LL (0LL), the Coulomb interaction gives rise to a particular SU(4) quantum Hall ferromagnet (QHFM) because of the fourfold spin-valley invariance symmetry in graphene monolayer. Then the spin and valley anisotropic energy terms explicitly break the SU(4) symmetry and lead to a diversity of QHFM phases, with distinct broken spin and valley symmetry[1–8]. Although previous studies have achieved great success in exploring the QHFM[9-12,17-22], the nature of the ground states at charge-neutral graphene, which is determined by a delicate balance between the spin and valley anisotropic energy terms, has remained under intense debate.

In the 0LL of graphene, the wavefunctions of each valley are locked to one of the real-space sublattices, which enables us to precisely pinpoint the nature of the QHFM phases by directly imaging the atomic-scale wavefunctions in real space, as first demonstrated in ref. 19 and further confirmed very recently in refs. 21 and 22. In this work, by atomic resolution imaging of electronic wavefunctions of charge-neutral graphene on insulating tungsten diselenide ($WSe_2$), we report microscopic signatures of a field-tuned continuous quantum phase transition of different valley orderings in the QHFM. During the phase transition, we observed field-dependent valley polarization and inversion, which are closely related to the observed anomalous field-dependent energy gap between 11.5 T and 14 T in the half-filled 0LL. Our work demonstrates the power to explore the nature and phase transitions of many-body orders by real-space atomic-scale imaging of the wavefunctions. The unexpected results obtained in this work, as well as the complex orders reported in literature[9-11,19-22], indicate that the true phase diagram of the QHFM at charge-neutral graphene is likely sample or substrate dependent, maybe partially arising from their different field-dependent spin and valley anisotropic energy terms, and needs further investigation.

In our experiment, the graphene/WSe$_2$ heterostructure was obtained by using a wet transfer fabrication of a single-crystal monolayer graphene on mechanical-exfoliated thick WSe$_2$ sheets (see methods for details of the sample preparation)[27]. Figure 1a shows a typical scanning tunneling microscope (STM) image of the graphene monolayer on multilayer WSe$_2$ substrate and the inset shows its corresponding fast Fourier transforms (FFT) image. We can observe moiré superlattice structure due to the lattice mismatch between WSe$_2$ and graphene (lattice constant: 0.353 nm for WSe$_2$ and 0.246 nm for graphene). Our experiment indicates that the Dirac point of graphene in the graphene/WSe$_2$ heterostructure depends on the thickness of the WSe$_2$ and the graphene can be changed from p-type doping to n-type doping by the thickness of the WSe$_2$ (see Fig. S1). Therefore, we can study electronic properties of charge-neutral graphene by carefully choosing the thickness of the supporting WSe$_2$. Figure 1b shows scanning tunneling spectroscopy (STS) measurements of two representative graphene/WSe$_2$ heterostructures, labeled as device 1 and 2 respectively, in a perpendicular magnetic field $B$ = 10 T. The tunneling spectra of both devices exhibit well-defined landau quantization of massless Dirac fermions (see Fig. S2), as expected in graphene monolayer. The 0LL of device 2 is about 35 meV above the Fermi level and that of device 1 is half filled. A pronounced feature is the splitting of the half-filled 0LL with a gap $\Delta E \approx 24$ meV at the Fermi level in device 1, which is indicative of a Coulomb gap when flat band or (quasi)bound state is partially filled[19,21,22,28-33]. The observed gap agrees quite well with the on-site Coulomb energy $E_C = e^2/4\pi\varepsilon_0\varepsilon_r l_B \sim$ 25 meV, where $e$ is the electron charge, $\varepsilon_0$ is the vacuum dielectric constant, $\varepsilon_r = (1+\varepsilon_{WSe_2})/2 \approx 4.4$ is relative dielectric constant surrounding the graphene ($\varepsilon_{WSe_2} \approx 7.8$ is the relative dielectric constant of WSe$_2$[34,35]), and $l_B = \sqrt{\hbar/eB}$ is the magnetic length with $\hbar$ the reduced Planck constant. Such a result indicates that the Coulomb interaction lifts the degeneracy of the half-filled 0LL and the gain in the exchange energy may favor the formation of spin/valley polarized/coherent orders in the charge-neutral graphene.

To further study the Coulomb gap of the 0LL, we measured high-resolution tunneling spectra in different magnetic fields in device 1 (see method for details). Figure 1c shows the 0LL as a function of the magnetic fields $B$. In small magnetic field region ($B < 8.2$ T), the 0LL is completely filled (with the filling factor as 2) and there is no Coulomb gap, further confirming that the splitting of the half-filled 0LL arises from the Coulomb interaction. The position of the 0LL depends strongly on the magnetic fields for $B < 8.2$

T, which arises from competition between magnetic confinement and spatial confinement generated by tip-induced electrostatic potential[36]. For the case that $B > 8.2$ T, the 0LL is half filled and the position of the 0LL is independent of magnetic fields, indicating that the magnetic length $l_B$ is much smaller than the characteristic length of the tip-induced potential and the effect of the tip gating is negligible[36]. Then, the observed gap at half-filled 0LL should be described by the on-site Coulomb energy. Figure 1d summarizes field dependent splitting of the 0LL obtained in device 1. The filling factor is 0 for $B > 8.2$ T and the splitting of the 0LL for $8.2$ T $< B < 11.5$ T agrees quite well with the on-site Coulomb energy $E_C = e^2 / 4\pi\varepsilon_0\varepsilon_r l_B$ of the graphene monolayer on WSe$_2$ substrate. However, the observed gap for the half-filled 0LL obviously deviates from the on-site Coulomb energy for $B > 11.5$ T. For simplicity, the observed charge gap in the 0LL of graphene should be the sum of the exchange (Coulomb) term and the lattice-scale symmetry-breaking terms, both of which can be strongly modified by the strength and/or direction of magnetic fields[4-7]. Therefore, the observed anomalous field-dependent charge gap in the half-filled 0LL for $B > 11.5$ T, as shown in Fig. 1d, suggests field-induced phase transitions between distinct lattice-scale orders in graphene.

To further explore origin of the anomalous field-dependent charge gap in the half-filled 0LL, we perform STS maps of the occupied and empty states of the 0LL, i.e., 0LL$^-$ and 0LL$^+$ respectively, at atomic scale. Although we observe well-defined hexagonal honeycomb lattices at Dirac point of graphene in zero magnetic field (Fig. S3), the STS maps of the occupied and empty states of the 0LL at $B = 10$ T exhibit a triangle lattice density distribution (Fig. 2a, left panels). Such a phenomenon indicates that the ground state of the charge-neutral graphene on $WSe_2$ at 10 T is the valley-polarized charge density wave (CDW) phase according to the sublattice-valley locking in the 0LL of graphene. There are two scenarios for the CDW phase as the ground state of the QHFM in charge-neutral graphene. In scenario one, the Coulomb interaction is strongly screened and the observed splitting of the 0LL is much smaller than the on-site Coulomb energy. Then, it is expected to observe the CDW phase in charge-neutral graphene, as reported very recently in graphene on a high dielectric constant substrate[21]. In our experiment, the observed charge gap of the 0LL agrees quite well with the on-site Coulomb energy. Therefore, we can rule out the scenario one. In scenario two, a moiré superlattice can add some sublattice symmetry breaking in graphene to stabilize the CDW phase, as observed in graphene on hexagonal boron nitride (hBN) with moiré superlattices[22]. In our device, the moiré superlattice structure generated between $WSe_2$ and graphene may help to stabilize the CDW phase at charge-neutral graphene for 8.2 T < $B$ < 11.5 T.

Along with the anomalous field-dependent charge gap in the half-filled 0LL for $B$ > 11.5 T, our detailed measurements of the ground states as a function of the magnetic field reveals a quantum phase transition with continuous changed valley polarization and, even more unexpected, valley inversion (Figure 2). At 10 T, the $0LL^-$ ($0LL^+$) is mainly polarized in the A (B) sublattice, as shown in Fig. 2a (left panels). However, the $0LL^-$ ($0LL^+$) becomes mainly polarized in the B (A) sublattice at 12.2 T (Fig. 2a, middle panels). By further increasing the field to 13.5 T, the $0LL^-$ ($0LL^+$) changes back to localize mainly in the A (B) sublattice (Fig. 2a, right panels). Figure 2b shows representative line cuts of the conductance maps measured at different magnetic fields (see more experimental results in Fig. S4). The changes of sublattice polarization with magnetic fields for $B > 11.5$ T indicate the field-induced valley inversion due to the sublattice-valley locking in the 0LL of graphene. To quantitatively analyze the field dependent sublattice (valley) polarization, we plotted $Z = \dfrac{I_A - I_B}{I_A + I_B}$ as a function of magnetic fields in Fig. 2c (here $I_A$ and $I_B$ are the intensity of conductance maps at the A and B sublattice respectively, as shown in Fig. 2b). The magnitude of $Z$ reflects the degree of sublattice (valley) polarization. For $Z = \pm 1$, we obtain fully valley polarized phase. In contrast, when $Z = 0$, the valley is completely un-polarized, indicating the spin is polarized at charge-neutral graphene[1–8]. For 8.2 T $< B <$ 11.5 T, we observe the valley-polarized CDW phase, however, the measured valley polarization is less than 1. Such a discrepancy is mainly attributed to the overlap of the 0LL$^-$ and 0LL$^+$, which are separated by about 20-30 meV and have a linewidth of about 20-30 meV, as shown in Figs. 1b and 1c. Therefore, we always observe a nonzero value on the A (B) sites in the conductance map of the 0LL$^+$ (0LL$^-$), even though it is predicted to be zero in the valley-polarized CDW phase. By increasing the fields to $B > 11.5$ T, the valley polarization decreases and reverses its direction at $B = 12.2$ T. The inverse valley polarization reaches its maximum at about 12.8 T and decreases to zero agrain at about 13.2 T. By further increasing the fields to $B > 13.2$ T, the valley polarization reverses back to its initial direction. After reaching the maximum at about 13.6 T, the valley polarization decreases to almost zero at $B = 14$ T. Obviously, our measurement reveals a quantum phase transition with continuous changed valley polarization and valley inversion for $B > 11.5$ T, which enriches the phase diagram of QHFM beyond current theory predictions. The observed phase transition may affect the lattice-scale symmetry-breaking terms and result in the anomalous field-dependent charge gap in the half-filled 0LL for $B > 11.5$ T.

In the graphene/WSe$_2$ heterostructure, edge atoms of WSe$_2$ nanostructures at the interface between graphene and WSe$_2$ could locally introduce electrostatic potential to change the doping of graphene[27], which allows us to study the effect of doping on the splitting of the 0LL locally. Figure 3a shows a representative STM image of graphene/WSe$_2$ heterostructure with a nanoscale WSe$_2$ quantum dot (QD) at the interface (see Fig. 3b for the schematic image of the structure). The WSe$_2$ QD generates a smooth electrostatic potential to bend the LLs and lifts the orbital degeneracy of the LLs around the WSe$_2$ QD, as shown in Fig. 3c. There is no intervalley scattering in graphene around the WSe$_2$ QD because that the electrostatic potential is quite smooth. For graphene above the nanoscale WSe$_2$ QD, there are both electron whispering-gallery modes and atomic collapse states due to the confinement of the local electrostatic potential (see Fig. S5 for detailed discussion)[27]. The bending of the LLs changes the doping of graphene and completely removes the splitting of the 0LL when the filling factor is changed from 0 to -2 (the 0LL is changed from half filling to complete empty), as shown in Figs. 3c and 3d. Such a result further confirms the Coulomb interaction as the origin of the splitting of the half-filled 0LL.

The bending of the LLs will generate topological conducting channel in graphene around the WSe$_2$ QD, as the topological edge states in the quantum Hall effect. Figure 4 shows STS maps in graphene around the WSe$_2$ QD at different energies. At the Fermi energy ($E$ = 0 meV), a ring-shaped structure is observed in graphene around the WSe$_2$ QD because the 0LL across the Fermi level (Fig. 4a). The width of such a conducting channel is about 5~10 nm and is quite robust when the bended 0LL is across the measured energy level, as shown in Figs. 4a and 4b. For the STS maps measured at 11.9 meV and -22.4 meV, we can also obtain the distribution of the 0LL$^-$ and 0LL$^+$ in graphene around the WSe$_2$ QD (Figs. 4b and 4c). For STS maps at energies between the (bended) LLs, we obtain uniform insulating feature in graphene around the WSe$_2$ QD, as shown in Fig. 4d as an example. Our result demonstrates that the local electrostatic potential can generate quantum Hall conducting channel inside a quantum Hall insulator, which is directly imaged by using STM. Such a result indicates that it is possible to study the quantum interference and excitation of the quantum Hall conducting channels by using STM.

**Acknowledgements**

This work was supported by the National Natural Science Foundation of China (Grant Nos. 11974050, 11674029) and National Key R and D Program of China (Grant No. 2021YFA1400100). L.H. also acknowledges support from the National Program for Support of Top-notch Young Professionals, support from "the Fundamental Research Funds for the Central Universities", and support from "Chang Jiang Scholars Program".

**Author contributions**

L. H. designed the experiment. Q. Z. synthesized the samples and performed the STM experiments. Q. Z. and L.H. analyzed the data and wrote the paper. All authors participated in the data discussion.

**Data availability statement**

All data supporting the findings of this study are available from the corresponding author upon reasonable request.

## Methods

**CVD Growth of Graphene**

The large area graphene monolayer films were grown on a $20\times20$ mm$^2$ polycrystalline copper (Cu) foil (Alfa Aesar, 99.8% purity, 25 μm thick) via a low pressure chemical vapor deposition (LPCVD) method. The cleaned Cu foil was loaded into one quartz boat in center of the tube furnace. Ar flow of 50 sccm (Standard Cubic Centimeter per Minutes) and H$_2$ flow of 50 sccm were maintained throughout the whole growth process. The Cu foil was heated from room temperature to 1030 ºC in 30 min and annealed at 1030 ºC for six hours. Then CH$_4$ flow of 5 sccm was introduced for 20 min to grow high-quality large area graphene monolayer. Finally, the furnace was cooled down naturally to room temperature.

**Construction of graphene/WSe$_2$ heterostructure**

We used conventional wet etching technique with polymethyl methacrylate

(PMMA) to transfer graphene monolayer onto the substrate. PMMA was first uniformly coated on Cu foil with graphene monolayer. We transferred the Cu/graphene/PMMA film into ammonium persulfate solution, and then the underlying Cu foil was etched away. The graphene/PMMA film was cleaned in deionized water for hours. The $WSe_2$ crystal was separated into thick-layer $WSe_2$ sheets by traditional mechanical exfoliation technology and then transferred to highly n-doped Si wafer [(100) oriented, 500 μm thick]. We placed graphene/PMMA onto Si wafer which has been transferred with $WSe_2$ sheets in advance. Finally, the PMMA was removed by acetone and then annealed in low pressure with Ar flow of 50 sccm and $H_2$ flow of 50 sccm at ~300 °C for 1 hours.

**STM and STS Measurements**

STM/STS measurements were performed in low-temperature (77 K and 4.2 K) and ultrahigh-vacuum (~$10^{-10}$ Torr) scanning probe microscopes [USM-1400 (77 K) and USM-1300 (4.2 K)] from UNISOKU. The tips were obtained by chemical etching from a Pt/Ir (80:20%) alloy wire. The differential conductance (d$I$/d$V$) measurements were taken by a standard lock-in technique with an ac bias modulation of 5 mV and 793 Hz signal added to the tunneling bias. In order to acquire the spectra with higher resolution, we lower the oscillation of lock-in bias voltage to 0.5 mV by using 1/10 bias divider during the STS measurement, where the energy resolution of high-resolution spectra is improved to about 1 meV.

# Figures

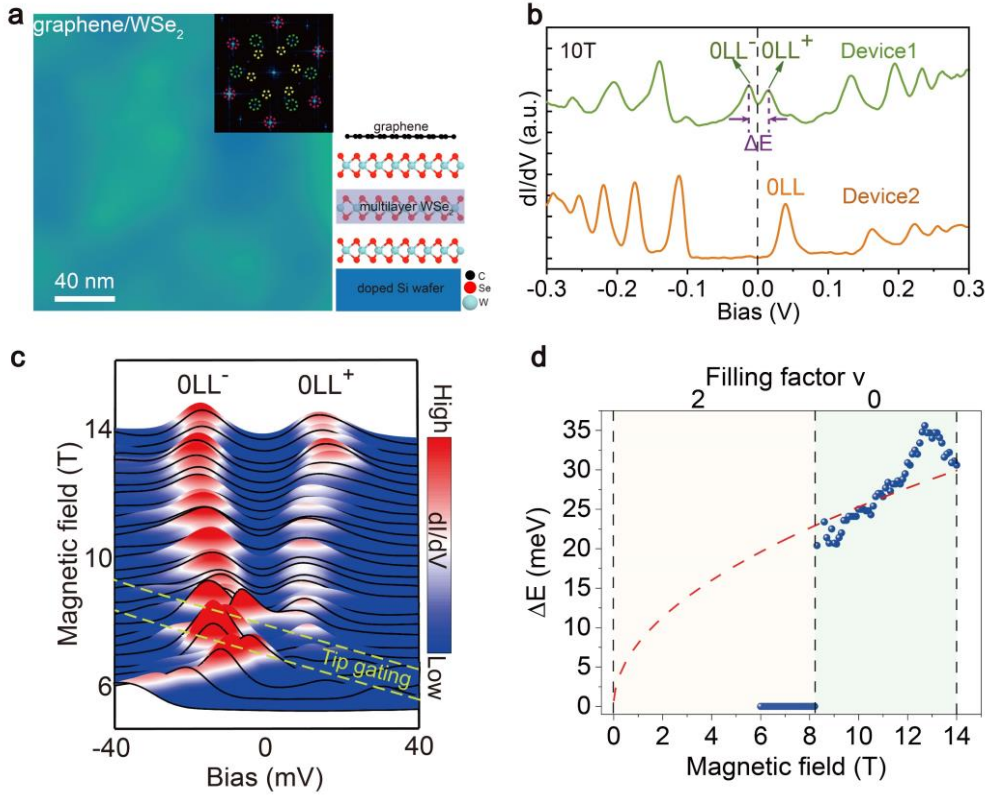

**Fig. 1 | The structure of Graphene/WSe$_2$ heterostructure and Coulomb gap of charge-neutral graphene. a,** Left: A STM image (bias $V$ = 600 mV, set point $I$ = 100 pA) of the graphene monolayer on WSe$_2$ substrate. Inset: the FFT of graphene/WSe$_2$ heterostructure. The spots in the red dotted circles represent the reciprocal lattice of graphene, the spots in the green dotted circles represent the reciprocal lattice of WSe$_2$, and the spots in the yellow dotted circles represent moiré structure of the graphene/WSe$_2$ heterostructure. Right: Schematic structure of the graphene/WSe$_2$ heterostructure. **b,** The typical d$I$/d$V$ spectra at $B$ = 10 T in the charge-neutral graphene (light green curve, device 1) and p-doped graphene (orange curve, device 2), respectively. In charge-neutral graphene, there is a splitting with about $\Delta E \approx 24$ meV in the 0LL. **c,** Magnetic fields dependent of the 0LL of the graphene in device 1. **d,** The energy separations of the two split peaks (acquired in panel **c**) as a function of magnetic fields $B$ (at the filling factor $\nu$ = 2 or 0). The dashed red line is the fitting result using the $\Delta E = e^2 / 4\pi\varepsilon_0\varepsilon_r l_B$ with $\varepsilon_{\mathrm{WSe}_2} \approx 7.8$.

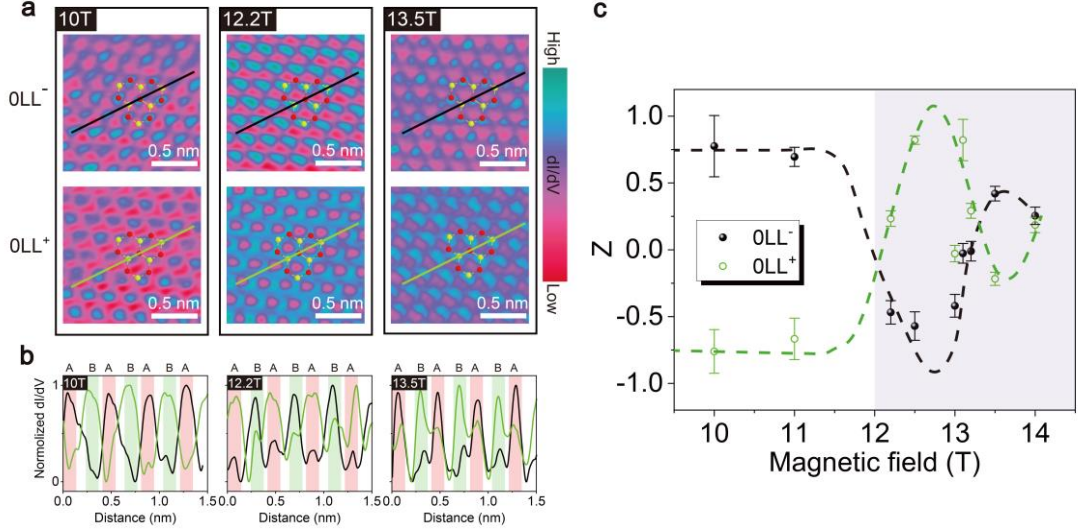

**Fig. 2 | Atomic-scale spatial distributions of the half-filled 0LL. a,** The STS maps of the 0 LL$^-$ and 0 LL$^+$ states at $v = 0$ under three typical magnetic fields. The hexagonal honeycomb structures of graphene are overlaid onto the STS maps. At $B = 10$ T, the sublattice polarized CDW phase is observed in the STS maps and the LDOS of two split peaks are localized in different graphene sublattices. However, at $B = 12.2$ T and 13.5 T, the d$I$/d$V$ maps reveal an anomalous phase transition with unexpected valley polarization and inversion from the CDW phase. **b,** The normalized intensity of STS signals along the black line (0 LL$^-$) and green line (0 LL$^+$) in panel **a**, showing intensity in the sublattice A (marked by red ball in panel **a**) and B (marked by green ball in panel **a**). **c**, The sublattice polarization $Z = \dfrac{I_A - I_B}{I_A + I_B}$ versus the magnetic field, where $I_A$ and $I_B$ are the intensity of d$I$/d$V$ signals at the A and B sublattice (acquired from panel **b**) respectively.

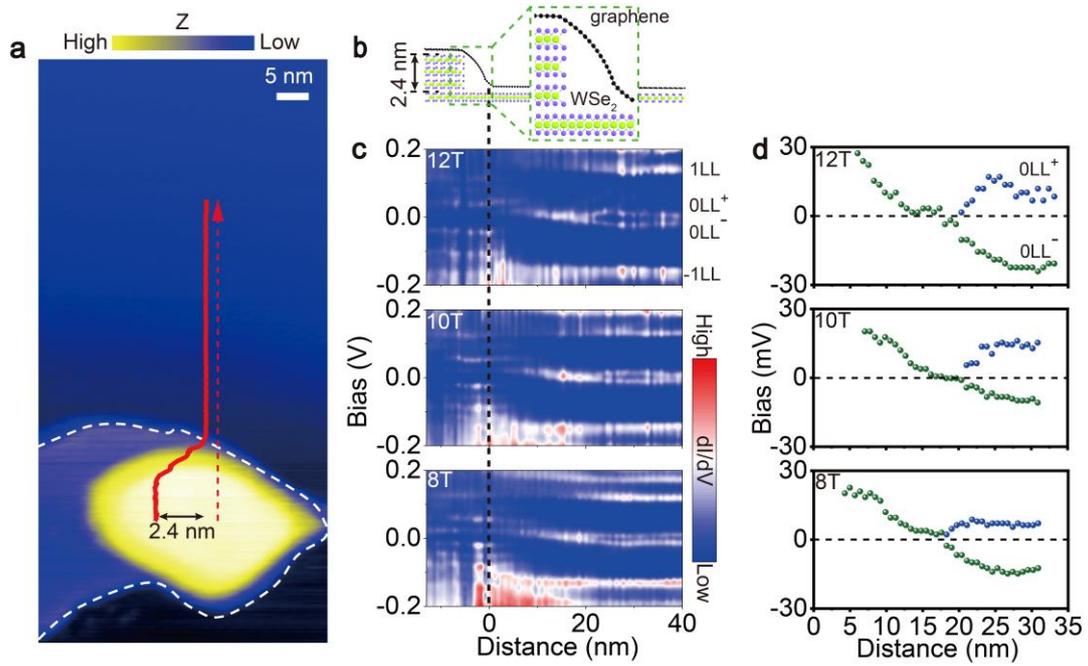

**Fig. 3 | Effects of local electrostatic potential on the LLs. a,** A representative STM image (bias $V = 600$ mV, set point $I = 100$ pA) of the graphene monolayer on $WSe_2$ substrate with an interface $WSe_2$ QD. The red solid line shows height profile along the dashed red arrow, indicating that the thickness of three-layer $WSe_2$ is about 2.4 nm. **b,** Schematic structure of the studied structure based on the STM image in panel **a**. **c,** STS spectroscopic maps along the dashed red arrow in the panel **a** under different magnetic fields ($B = 8, 10, 12$ T). The solid black line indicates the edge of the $WSe_2$ QD. **d,** The 0 LL$^-$ and 0 LL$^+$ states from panel **c** as a function of distance from the edge of the $WSe_2$ QD.

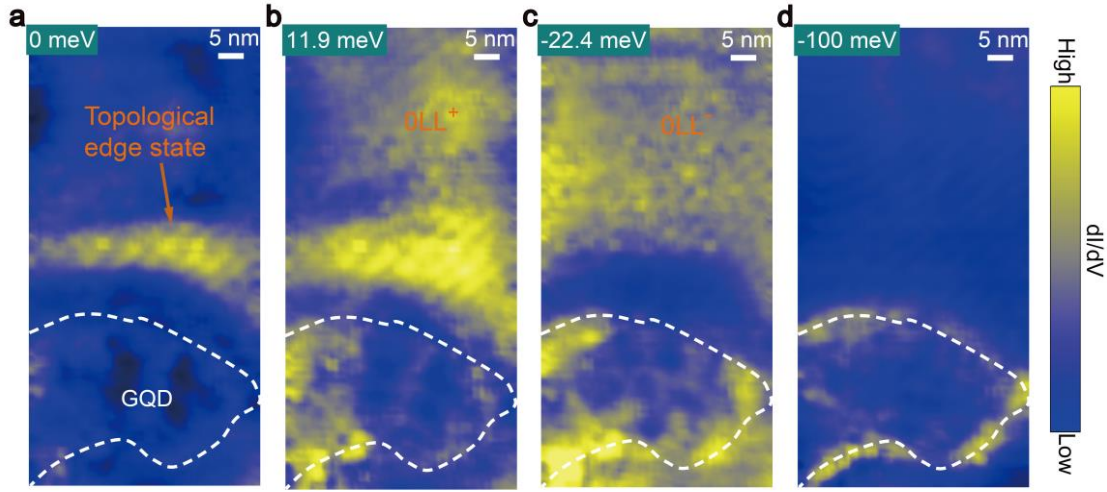

**Fig. 4 | Spatial distributions of electronic states at different energies in the QH region. a,** $E = 0$ meV. **b,** $E = 11.9$ meV. **c,** $E = -22.4$ meV. **d,** $E = -100$ meV. The magnetic field is 12 T. At $E = 0$ meV, a ring-shaped QH topological conducting channel can be observed in a region about 15 ~ 25 nm away from the edge of the WSe$_2$ QD due to the electrostatic potential. The maps at 11.9 meV and -22.4 meV show distributions of the 0LL$^+$ and 0LL$^-$ states around the WSe$_2$ QD. The map at -100 meV shows the insulating state between the 0LL and -1LL.